\documentclass[sigconf]{acmart}%,authordraft
\usepackage[ruled,linesnumbered,vlined]{algorithm2e}
\usepackage{multirow}
\usepackage{subfigure}
\usepackage{colortbl}

\AtBeginDocument{%
  \providecommand\BibTeX{{%
    \normalfont B\kern-0.5em{\scshape i\kern-0.25em b}\kern-0.8em\TeX}}}
\setcopyright{rightsretained}

\setcopyright{acmcopyright}
\copyrightyear{2022}
\acmYear{2022}
\acmDOI{XXXXXXX.XXXXXXX}

\acmConference[CIKM'22]{the 31st ACM International Conference on Information and Knowledge Management}{October 17-22, 2022}{Hybrid Conference, Hosted in Atlanta, Georgia, USA}
\acmPrice{15.00}
\acmISBN{978-1-4503-XXXX-X/18/06}

\begin{document}

\theoremstyle{definition}
\newtheorem{define}{Definition}[]

%%
%% The "title" command has an optional parameter,
%% allowing the author to define a "short title" to be used in page headers.
\title{From Known to Unknown: Quality-aware Self-improving Graph\\ Neural Network for Open Set Social Event Detection}

% \author{Jiaqian Ren$^{1,2}$, Lei Jiang$^2$, Hao Peng$^3$, Yuwei Cao$^4$, Jia Wu$^5$, Philip S. Yu$^4$, Lifang He$^6$}
% \affiliation{
% \institution{
% $^1$ Institute of Information Engineering, Chinese Academy of Sciences;\\
% $^2$School of Cyber Security, University of Chinese Academy of Sciences;\\
% $^3$Beihang University; $^4$University of Illinois Chicago;$^5$Macquarie University; $^6$Lehigh University.\\
% $^*$ Corresponding authors (Jianglei@iie.ac.cn, penghao@buaa.edu.cn)
% }
% }

\author{Jiaqian Ren}
\affiliation{%
   \institution{Institute of Information Engineering,\\ Chinese Academy of Sciences}
   \country{China}}
\email{renjiaqian@iie.ac.cn}

\author{Lei Jiang}
\affiliation{%
   \institution{Institute of Information Engineering,\\ Chinese Academy of Sciences}
   \country{China}}
\email{jianglei@iie.ac.cn}
\authornote{Corresponding authors}

\author{Hao Peng$^*$}
\affiliation{%
   \institution{Beihang University}
   \country{China}}
\email{penghao@buaa.edu.cn}

\author{Yuwei Cao}
\affiliation{%
   \institution{University of Illinois Chicago}
   \country{USA}}
\email{ycao43@uic.edu}

\author{Jia Wu}
\affiliation{%
   \institution{Macquarie University}
   \country{Australia}}
\email{jia.wu@mq.edu.au}

\author{Philip S. Yu}
\affiliation{%
   \institution{University of Illinois Chicago}
   \country{USA}}
\email{ psyu@uic.edu}

\author{Lifang He}
\affiliation{%
   \institution{Lehigh University}
   \country{USA}}
\email{lih319@lehigh.edu}
%\email{{renjiaqian, Jianglei}@iie.ac.cn, penghao@buaa.edu.cn, {ycao43, psyu}@uic.edu, jia.wu@mq.edu.au, lih319@lehigh.edu.}
%\thanks{Jiang Lei and Hao Peng are corresponding authors.}
% \affiliation{
% \institution{
% $^1$ Institute of Information Engineering, Chinese Academy of Sciences, Beijing 100093, China;\\
% $^2$School of Cyber Security, University of Chinese Academy of Sciences, Beijing 100049, China;\\
% $^3$School of Cyber Science and Technology, Beihang University, Beijing 100191, China;\\
% $^4$Department of Computer Science, University of Illinois Chicago, IL 60607, USA;\\
% $^5$Department of Computing, Macquarie University, Sydney NSW 2109, Australia;\\
% $^6$Department of Computer Science and Engineering, Lehigh University, Bethlehem, PA 18015, USA.\\
% }
% }
% \email{{renjiaqian, Jianglei}@iie.ac.cn, penghao@buaa.edu.cn, {ycao43, psyu}@uic.edu, jia.wu@mq.edu.au, lih319@lehigh.edu.}
% \thanks{Jiang Lei and Hao Peng are corresponding authors.}

\renewcommand{\shortauthors}{Jiaqian Ren, et al.}

\begin{abstract}
State-of-the-art Graph Neural Networks (GNNs) have achieved tremendous success in social event detection tasks when restricted to a closed set of events. However, considering the large amount of data needed for training a neural network and the limited ability of a neural network in handling previously unknown data, it remains a challenge for existing GNN-based methods to operate in an open set setting. To address this problem, we design a 
Quality-aware Self-improving Graph Neural Network (QSGNN) which extends the knowledge from known to unknown by leveraging the best of known samples and reliable knowledge transfer. Specifically, to fully exploit the labeled data, we propose a novel supervised pairwise loss with an additional orthogonal inter-class relation constraint to train the backbone GNN encoder. The learnt, already-known events further serve as strong reference bases for the unknown ones, which greatly prompts knowledge acquisition and transfer. When the model is generalized to unknown data, to ensure the effectiveness and reliability, 
we further leverage the reference similarity distribution vectors for pseudo pairwise label generation, selection and quality assessment. Following the diversity principle of active learning, our method selects diverse pair samples with the generated pseudo labels to fine-tune the GNN encoder. Besides, we propose a novel quality-guided optimization in which the contributions of pseudo labels are weighted based on consistency. We thoroughly evaluate our model on two large real-world social event datasets. Experiments demonstrate that our model achieves state-of-the-art results and extends well to unknown events. 
\end{abstract}

\begin{CCSXML}
<ccs2012>
 <concept>
  <concept_id>10010520.10010553.10010562</concept_id>
  <concept_desc>Computer systems organization~Embedded systems</concept_desc>
  <concept_significance>500</concept_significance>
 </concept>
 <concept>
  <concept_id>10010520.10010575.10010755</concept_id>
  <concept_desc>Computer systems organization~Redundancy</concept_desc>
  <concept_significance>300</concept_significance>
 </concept>
 <concept>
  <concept_id>10010520.10010553.10010554</concept_id>
  <concept_desc>Computer systems organization~Robotics</concept_desc>
  <concept_significance>100</concept_significance>
 </concept>
 <concept>
  <concept_id>10003033.10003083.10003095</concept_id>
  <concept_desc>Networks~Network reliability</concept_desc>
  <concept_significance>100</concept_significance>
 </concept>
 <concept>
<concept_id>10010147.10010178</concept_id>
<concept_desc>Computing methodologies~Artificial intelligence</concept_desc>
<concept_significance>500</concept_significance>
</concept>
<concept>
<concept_id>10002951.10003227.10003351</concept_id>
<concept_desc>Information systems~Data mining</concept_desc>
<concept_significance>500</concept_significance>
</concept>
</ccs2012>
\end{CCSXML}

\ccsdesc[500]{Computing methodologies~Artificial intelligence}
\ccsdesc[500]{Information systems~Data mining}

%%
%% Keywords. The author(s) should pick words that accurately describe
%% the work being presented. Separate the keywords with commas.
\keywords{Social event detection, graph neural network, contrastive learning, active learning}

%% This command processes the author and affiliation and title
%% information and builds the first part of the formatted document.
\maketitle
\section{Introduction}
The goal of social event detection is to discover meaningful events from the social media data by grouping messages reported on the same event together~\cite{atefeh2015survey,li2021deep}. Due to its broad range of potential applications such as business marketing, disaster risk management, public opinion analysis, etc., social event detection has experienced explosive growth in recent years~\cite{fedoryszak2019real}. 

Previous studies often utilize text contents~\cite{zhao2011comparing,yan2015probabilistic,amiri2016short,wang2017neural,sahnoun2020event,morabia2019sedtwik} or auxiliary attributes extracted from texts~\cite{xie2016topicsketch,xing2016hashtag,feng2015streamcube} to make social event detection. However, in recent years, there is a trend towards GNN-based methods~\cite{peng2019fine,cao2021knowledge,peng2021streaming,peng2022reinforced,cui2021mvgan} which have the ability to combine contents and attributes together. By fully capturing the rich semantics and structural 
information contained in the social data, GNN-based approaches achieve remarkable performance when restricted to a closed set of events. Taking a deep dive into these neural network models, their success heavily relies on the huge amount of data with human annotations under the closed-world assumption. However, the data of social networks is being updated all the time, which means there are lots of newly emerging events. On the one hand, consistently 
identifying and annotating all emerging categories for model training are cost-inhibitive. On the other hand, given the presence of significant distribution gaps between the already-known training events and the emerging unknown events, directly applying the models trained on known data to unknown data in the real world typically exhibits clear performance degradation. When the conventional closed set assumption no longer holds, how to keep the strength of GNN methods as well as generalize them to open set application is a challenging problem.

We argue that the aforementioned problem can be solved by human-like learning processes. A human can easily identify samples of new events after they are trained to distinguish events with some already-known samples. Likewise, for the training of models, it is natural to leverage the already-known events (the annotated data) to explore the new, unlabeled events, rather than in a completely unsupervised way. Generally, the success of human beings in recognizing new events is due to two kinds of abilities: 1) the ability to find the pattern of an event with already-known samples, and 2) the ability to transfer knowledge from known to unknown. Analogously, the key challenges in open set social event detection are: 1) how to fully exploit the labeled data to learn discriminative features, and 2) how to achieve effective and reliable knowledge transfer to promote the discovery of new events. However, existing GNN-based event detection methods still have room for improvement on the first issue, and totally ignore the second one. Consider the first challenge, authors in \cite{peng2019fine,cui2021mvgan} train GNN networks to learn the discriminative features of events based on the cross-entropy loss. While great results have been achieved, they are unsuitable for the open world. Authors in \cite{cao2021knowledge} and \cite{peng2022reinforced} replace the cross-entropy loss with 
a triplet loss to train the model. However, they still cannot fully exploit data. This is because the complicated sampling strategy in the triplet loss brings a weaker generalization capability when being extended to the unknown set, and the relative distances between positive and negative pairs learnt from the triplet loss are not always distinguishable (the intra-class distances are sometimes larger than the inter-class distances). To solve these problems, we propose a stricter pairwise loss in this paper. 

% Our pairwise loss holds the idea that the distance between an intra-class pair should be smaller than that of any inter-class pairs no matter whether they share an anchor or not.
\begin{figure}
\centering
\includegraphics[width=0.47\textwidth]{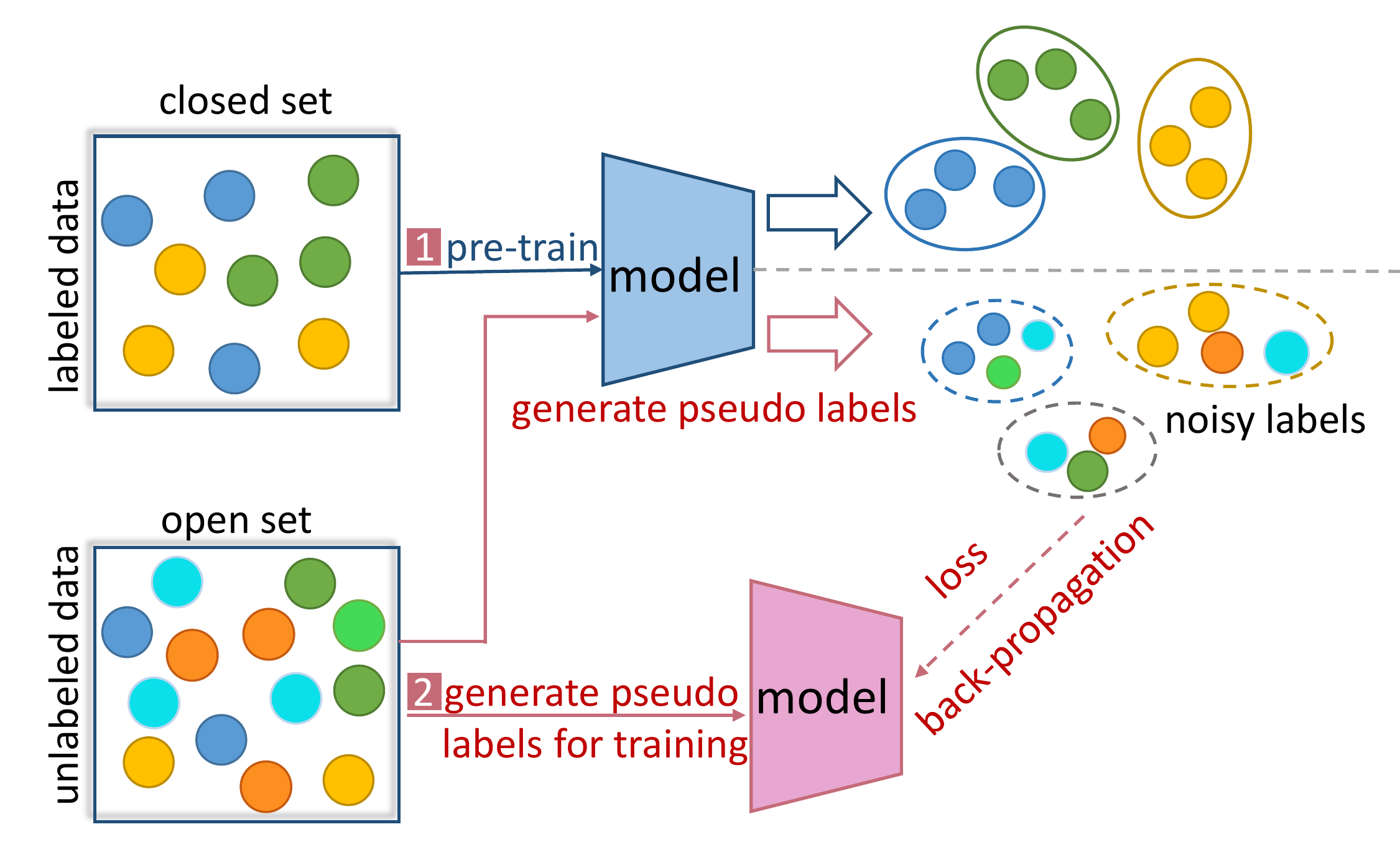}
\caption{The two-step learning strategy for knowledge transferring from known to unknown.
} \label{fig_trans}
\end{figure}

As to the second challenge, i.e., knowledge transferring from known to unknown, those GNN-based methods~\cite{cao2021knowledge,peng2022reinforced} which consider the incremental, novel events setting in the real world, need labeled data for continuous training. However, constantly labeling new data is time-consuming and labor-intensive. To achieve knowledge transfer in an unsupervised way, it is worth noting that a series of methods~\cite{hsu2018multi,han2019learning,han2020automatically,zhong2021openmix,jia2021joint,zhong2021neighborhood} are proposed for the task of novel class discovery. As illustrated in Fig.~\ref{fig_trans}, these methods commonly follow a two-step learning strategy: 1) pre-training the model with labeled data to obtain basic discriminative ability; 2) adopting clustering or pairwise determination algorithms to generate pseudo labels for the training of the new classes. 
However, using all pseudo labels for training is unnecessary and time-consuming. Besides, these methods fail to consider the noise in the obtained pseudo-labels and, therefore, are unreliable.
How to select a small number of high-quality samples to achieve effective and reliable knowledge transfer becomes an important problem. 
In a word, the task of open set social event detection is quite hard and has not been well-solved yet.

To tackle the above challenges, in this work, we design a Quality-aware Self-improving Graph Neural Network (QSGNN) which extends the knowledge from known to unknown by leveraging the best of the known samples and reliable knowledge transfer. Specifically, to fully explore the discriminative knowledge from the already-known data, we propose a novel pairwise learning method which has a stricter demand on the distances between intra-class samples and inter-class ones to train our backbone GNN encoder. In addition to differentiating the known events by the distances in the latent space, we also add an orthogonal inter-class constraint to force the known events to scatter in different directions. In this way, those known events are fully explored and become strong reference bases for unknown events. To ensure the effectiveness and reliability of the knowledge transferring process, we propose to utilize the Reference Similarity Distribution (RSD) vector, which is obtained by computing the similarity distribution over a set of known reference events, for pseudo pairwise label generation, selection and quality assessment. Particularly, we follow the principle of diversity in active learning~\cite{ren2021survey} to make unbalanced sample selection, and assess the quality of the pseudo labels by their consistencies. To further resist the noise of the pseudo labels, we adopt a quality-guided optimization in which the contributions of the the pseudo labels are weighted. We continuously fine-tune the backbone GNN encoder to adapt to the incremental, unseen events in the open world. Note that our model provides a principled way to obtain pairwise pseudo labels with high quality for unlabelled data, thus enabling effective and reliable knowledge transfer.

Experimental results on two large and publicly available Twitter datasets show that our method achieves state-of-the-art performance in closed set event detection and maintains high performance in the open set setting.
The source code and data are available at GitHub\footnote{\url{https://github.com/RingBDStack/open-set-social-event-detection}}.
Our main contributions are summarized as follows: 
1) We propose a Quality-aware Self-improving Graph Neural Network (QSGNN) which extends knowledge from known to unknown by leveraging the best of known
samples and reliable knowledge transfer. It successfully solves the open set social event detection problem.
2) We propose a novel pairwise learning method with an orthogonal inter-class constraint. Our method demands stricter distance distinctions as well as direction distinctions in the latent space, thus fully exploits the knowledge contained in the labeled data. 3) We ensure effective and reliable knowledge transfer from known to unknown by a selection strategy based on diversity as well as a quality assessment strategy based on consistency. The effectiveness comes from the unbalanced sample selection, and the reliability is guaranteed by the quality-guided optimization process.
% \begin{itemize}
% \item We propose a Quality-aware Self-improving Graph Neural Network (QSGNN) which extends the knowledge from known to unknown by leveraging the best of known
% samples and reliable knowledge transferring. Our model successfully solves the open set social event detection problem.
% \item We propose a novel pairwise contrastive learning method with an orthogonal inter-class constraint. Our method demands more strict distance distinctions as well as direction distinctions in the latent space, thus fully exploiting the knowledge contained in the labeled data. \item We make effective and reliable knowledge transferring from known to unknown by a selection strategy based on diversity as well as a quality assessment strategy based on consistency. The effectiveness is provided by the unbalanced sample selection, and the reliability is provided by the quality-guided optimization process.
% \end{itemize}

% This module is important for both knowledge acquiring and reliable knowledge transferring. 
% Benefiting from the evidential theory, it 

% However, limited progress has been achieved for open set social event detection which is increasingly valuable in practice. 

\section{Related Work}
\begin{figure*}
\centering
\includegraphics[width=0.88\textwidth]{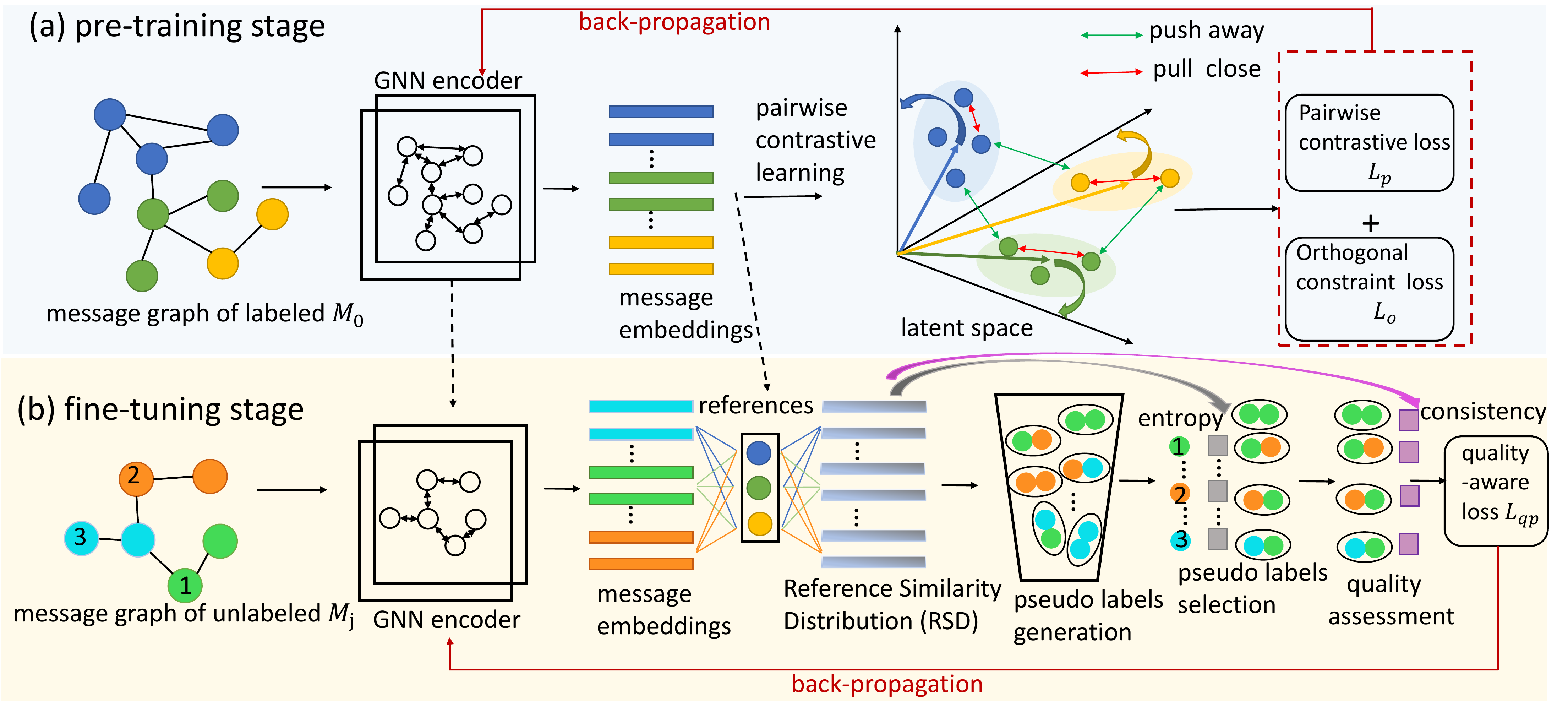}
\caption{The architecture of QSGNN. (a) shows the pre-training stage, in which we fully explore the discriminative knowledge from the labeled data by utilizing both distance and direction
information. 
%We propose pairwise contrastive learning with an orthogonal inter-class relation constraint to train our backbone GNN encoder. 
(b) demonstrates the fine-tuning stage, in which we utilize the Reference Similarity Distribution vector for pseudo pairwise label generation, selection and quality assessment. 
% We fine-tune the backbone model to adapt to the emerging unlabeled data. 
} \label{fig_model}
\end{figure*}
\subsection{Social Event Detection}
Event detection in social networks has received considerable attention in recent years~\cite{goswami2016survey}. 
Based on existing techniques, social event detection methods can be divided into incremental clustering ones~\cite{aggarwal2012event,ozdikis2017incremental}, community detection ones~\cite{fedoryszak2019real,liu2020story}, and topic modeling ones~\cite{zhao2011comparing,yan2015probabilistic,xie2016topicsketch,xing2016hashtag}. Though the incremental clustering methods can be easily adapted to open set social event detection tasks, they fail to fully explore the knowledge contained in the social streams as they ignore the rich semantics and structural information. This problem also exists in the community detection methods and the topic modelling methods.
Based on the information they exploit, social event detection methods can be categorized
into three types: content-based~\cite{zhao2011comparing,yan2015probabilistic,amiri2016short,wang2017neural,sahnoun2020event,morabia2019sedtwik}, attribute-based~\cite{xie2016topicsketch,xing2016hashtag,feng2015streamcube}, and combined methods~\cite{liu2020event,wang2016using,peng2019fine,cao2021knowledge,peng2021streaming,peng2022reinforced,cui2021mvgan}. Among the combined methods, the GNN-based ones~\cite{peng2019fine,cao2021knowledge,peng2021streaming,peng2022reinforced,cui2021mvgan,ren2022evidential,ren2021transferring} perform greatly due to their powerful expressive ability in building social graphs to effectively combine contents and attributes. However, most GNN-based studies hold the closed-set assumption in which the training set and test set share the same events thus cannot be directly applied in the open world. Few exceptions~\cite{cao2021knowledge,peng2021streaming,peng2022reinforced} do consider the incremental events setting but still need annotated data for continuous training, which is costly. Considering the limited ability of neural networks in handling previously unknown data, adapting GNN-based methods to the open world setting remains challenging.

\subsection{Active Learning}
Active learning, also known as query learning, assumes that different samples in the dataset have different values for the model training, and tries to select a small quantity of data to achieve high performance gains~\cite{ren2021survey}. Based on the query principle they employ, active learning methods can be split into three categories: uncertainty-based methods~\cite{tong2001support,beluch2018power,ranganathan2017deep,9492291}, diversity-based methods~\cite{bilgic2009link,gal2017deep,guo2010active,nguyen2004active,wu2021redal}, and expected model change-based ones~\cite{freytag2014selecting,roy2001toward,settles2007multiple,ijcai2018-296}. In this work we follow the principle of diversity and propose an unbalanced sampling strategy.

%Event detection over twitter social media streams
\subsection{Novel Class Discovery}
The task of novel class discovery is proposed recently aiming at recognizing novel classes in unlabeled data~\cite{han2019learning}, which is similar to our open set event detection task.
This task differs from the conventional unsupervised learning as they leverage some already-known data. Existing methods~\cite{hsu2018multi,han2019learning,han2020automatically,zhong2021openmix,jia2021joint,zhong2021neighborhood} usually follow a two-step strategy: 1) using the labeled data for model initialization, 2) performing unsupervised clustering or pairwise determination on unlabeled data to fine-tune the model. For example, to recognize open set images, authors in \cite{hsu2018multi} propose a constrained clustering network. They first measure the pairwise similarities between images by training a classification model on labeled data, then adopt a clustering model on unlabeled data with those pairwise predictions. Later, authors in ~\cite{han2020automatically} directly utilize rank statistics to estimate the pairwise similarity of images.
Though these approaches which use pseudo labels to achieve model adaption to the unlabelled data have achieved
promising results, they fail to consider the noise contained in the obtained pseudo-labels. Therefore, the training process is unreliable. What's more, they have not given a sample selection strategy. Considering the large amount of unlabeled data, how to prompt performance with a small number of samples becomes an important problem.
% Based on this finding, we also build on the premise of leveraging pseudo-label but propose a diversity-based selection strategy as well as a quality assessment strategy.

% \subsection{Contrasive Learning}

\section{Methodology}
In this section, we begin by presenting the problem definition of the open set social event detection task in Sec.~\ref{model_pd}, and then introduce the details of our model. Fig.~\ref{fig_model} shows an overview of our proposed QSGNN, which includes the supervised pre-training stage (Sec.~\ref{model_pretrain}) and the self-improving fine-tuning stage (Sec.~\ref{model_finetune}).

\subsection{Problem Definition}\label{model_pd}
Here we give the definition of the open set social event detection task in this work. The goal of our work is to identify newly emerging events in unlabelled incremental social streams with the support of knowledge learnt from the existing known events.

Formally, the open set event data comes as incremental social stream~\cite{cao2021knowledge}. The social stream, denoted as $S=M_0,M_1,...,M_{i-1},M_{i}$, is actually a temporal sequence of blocks of social messages, where each message block $M_j, j\in\{0,1,...,i\}$ contains all the messages that arrive during the split time period. Different from \cite{cao2021knowledge}, in this work we impose the constraint that only the messages in the initial block (i.e., $M_0$) are provided with labels. The later blocks remain unlabeled during the whole training process. The objective is two-fold: 1) In the pre-training stage, from the labeled message block $M_0$, we learn an initial detection encoder, $f(M_0,\theta_0)$, that extracts discriminative features and detects events. $\theta_0$ denotes the parameter of $f(M_0,\theta_0)$ and is trained from $M_0$. 2) To detect events from the unlabeled message blocks, we extend knowledge from known to unknown by consistently updating the detection encoder. That is, we learn a sequence of detection encoders $f(M_j,\theta_0,\theta_j), j\in\{1,...,i\}$. Each of them is fine-tuned on the corresponding unlabeled message block to achieve knowledge transferring.

\subsection{Overall Framework}\label{model_framework}
Our work contains two stages: 1) the model pre-training stage with supervised pairwise contrastive learning (Eq.~(\ref{equ_pairloss})), and 2) the model fine-tuning stage with quality-aware self-improving learning (Eq.~(\ref{equ_adj_pairloss})). 
% Considering numerous social messages appear in the social network, treating them as a whole requires a lot of storage and computing resources. 
As mentioned in Sec.~\ref{model_pd}, we split the whole dataset into an incremental social stream. In the pre-training stage, an initial message graph is constructed. We assume those messages are all labeled and utilize them to train our backbone GNN encoder. In the fine-tuning stage, messages are all assumed to be unlabeled. For each message block, we construct a new graph. We directly input each coming block to the already trained backbone model to get initial representations for pseudo labels generation, selection, and quality assessment. Note that for each unlabeled block, we consistently update the model with the generated pseudo labels and re-generate new pseudo labels for three times. In this way, the model adapts to the incoming data. Besides, utilizing only the current message block for fine-tuning, our procedure maintains a light training scheme.

\subsection{Supervised Pre-training of Backbone GNN Encoder}\label{model_pretrain}
\subsubsection{Construction of message graph}
Here we give a brief introduction to the message graph construction process. The message graph, which only contains message nodes, is used to express the complicated internal relations of messages in the social stream. To fully explore all kinds of important information contained in the social stream, we build edges between messages based on three types of elements including users, hashtags, and entities. Messages share any of these three elements are linked together.  
As for the initial message features, we follow the way in \cite{cao2021knowledge} which combines the language semantics with temporal information together. Specifically, the semantic feature of a message is obtained by averaging the pre-trained embeddings~\cite{mikolov2013} of its words. The 2-d temporal feature is obtained by converting the timestamp to the OLE date. The initial message feature is the concatenation of these two parts.

\subsubsection{Backbone GNN encoder}
After constructing the message graph, we apply a GNN encoder on it to learn comprehensive message representations. To fully incorporate the rich semantics and relations, the GNN encoder learns node representations by iteratively combining information from their one-hop neighbours. Formally, for message $m_i$, whose representation in the $l$-th layer is denoted as $\boldsymbol{h}_{m_i}^l$, its updated representation in the $(l+1)$-th layer becomes:
\begin{equation}
\boldsymbol{h}_{m_{i}}^{(l+1)} \leftarrow \stackrel{\text { heads }}{\|}\left(\boldsymbol{h}_{m_{i}}^{(l)} \oplus \underset{\forall m_{j} \in \mathcal{N}\left(m_{i}\right)}{\operatorname{Aggregator}}\left(\text { Extractor }\left(\boldsymbol{h}_{m_{j}}^{(l)}\right)\right)\right),
\end{equation}
where $\mathcal{N}\left(m_{i}\right)$ denotes the one-hop neighbours of message $m_{i}$. $\oplus$ stands for an aggregation, and $\stackrel{\text { heads }}{\|}$ represents concatenation of multiple heads. $\text {Aggregator}$ and $\text {Extractor}$ strategies differ in different GNN models. In this paper we adopt the strategies in GAT~\cite{velivckovic2018graph}.

\subsubsection{Supervised pairwise contrastive learning}
As new events continuously arrive in the open world, the total number of events is hard to know in advance. Thus, cross-entropy loss functions that are widely used in the closed set are not suitable anymore. To cope with the discrepancies between the closed and the open set settings, we need to design a training method which can perform well under supervision as well as be easily generalized to unknown samples. Accordingly, we propose a novel pairwise learning method, which is inspired by the triplet loss in \cite{schroff2015facenet}, but different in the removal of anchors. The essence of our pairwise learning is the idea that the distance between an intra-class pair should be smaller than any of the inter-class pairs, no matter whether they share a common node or not. 

In \cite{schroff2015facenet}, to compute the triplet loss, one must construct a set of triplets $\{T\}$ first. Each triplet is composed of an anchor, a positive sample to the anchor (i.e., a sample within the same class), and a negative sample to the anchor (i.e., a sample from a different class). Suppose the anchor message is $m_i$. $m_i+$ and $m_i-$ denote the positive and the negative samples, respectively. The triplet loss is as follows:
\begin{equation}
    \mathcal{L}_{t}=\sum_{\left(m_{i}, m_{i}+, m_{i}-\right) \in \{T\}} \max \left\{\mathcal{D}\left(\boldsymbol{h}_{m_{i}}, \boldsymbol{h}_{m_{i}+}\right)-\mathcal{D}\left(\boldsymbol{h}_{m_{i}}, \boldsymbol{h}_{m_{i}-}\right)+a, 0\right\},
\end{equation}
where $a$ is a hyper-parameter which represents the margin distance. $D(\cdot,\cdot)$ computes the Euclidean distance.

\begin{figure}
\centering
\includegraphics[width=0.43\textwidth]{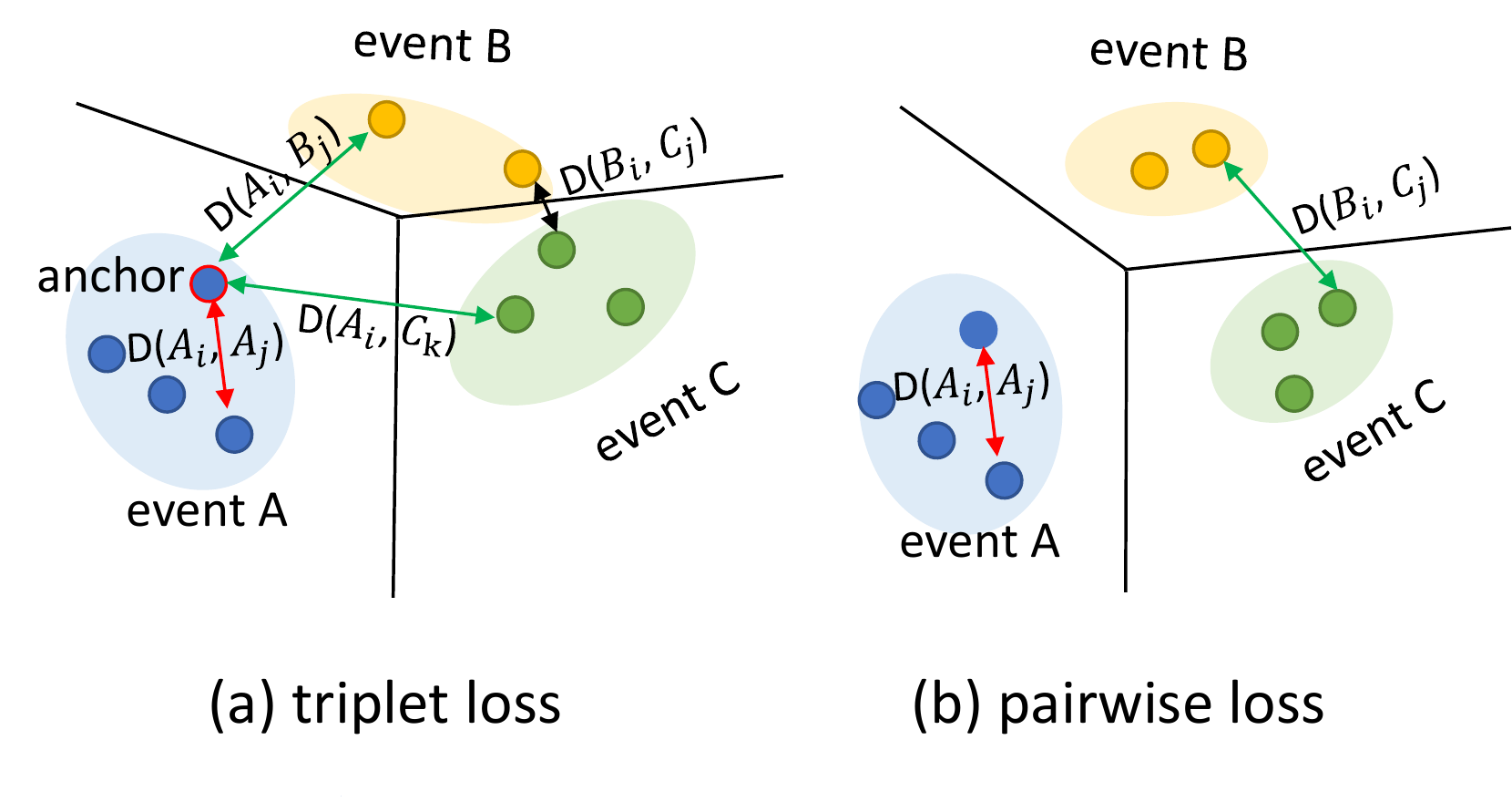}
\caption{(a) and (b) illustrate the representation distributions learnt by the triplet loss vs. the pairwise loss (ours). Our pairwise loss guarantees that the intra-class distances are always smaller than the inter-class ones, e.g., $D(A_i, A_j) < D(B_i, C_j)$. } \label{fig_pair}
\end{figure}

Carefully analysing the triplet loss computation process mentioned above, we find two deficiencies: 1) Its sampling strategy is complicated, which hampers its generalization from the known set to the unknown set. When the labels are provided, one can easily obtain triplets by constructing all possible positive pairs in a specific class and adding the same number of random negative samples from the other classes. However, for unknown data, the triplets are much harder to construct. Specifically, all node pairs need to be classified before we can combine the positive ones with other samples to form triplets. %However, the additional alignment does not needed in our pairwise learning.
%in our pairwise contrastive loss, with the removal of anchors, we only need a series of positive pairs and a series of negative pairs which are easy to sample and generalize.
2) The triplet loss which trains the model based on the relative distances between positive
and negative pairs w.r.t. the same anchor is a relaxed constraint. As shown in Fig.~\ref{fig_pair}(a), after training, the intra-class distance in event $A$ may be larger than the inter-class distance between event $B$ and event $C$. E.g., $D(B_i,C_j)$ is smaller than $D(A_i,A_j)$, which creates difficulties for the correct clustering of events.

To simplify the sampling process and make the intra-class distributions more distinguishable from the inter-class ones,
our pairwise loss introduces a stricter constraint which pushes away negative pairs from positive pairs despite the anchors. The loss is as follows:
\begin{equation}\label{equ_pairloss}
\mathcal{L}_{p}=\sum_{(m_{i}, m_{i}+)\in\{Pos\}
    \atop
    (m_j, m_{j-})\in\{Neg\} } \max \{\mathcal{D}(\boldsymbol{h}_{m_{i}}, \boldsymbol{h}_{m_{i}+})-\mathcal{D}(\boldsymbol{h}_{m_{j}}, \boldsymbol{h}_{m_{j}-})+a, 0\},
\end{equation}
where $\{Pos\}$ denotes the positive pairs and $\{Neg\}$ denotes the negative ones. Obviously, with the help of this loss, the minimum inter-class distance in the whole batch is required to be larger than the maximum intra-class distance, as shown in Fig.~\ref{fig_pair}(b). This helps differentiate the intra-class representations from the inter-class ones thus improves the events clustering results.

\subsubsection{Orthogonal inter-class relation constraint}
The pairwise loss distinguishes the events by solely controlling the distances between messages in the latent space. There is no utilization of the direction information. Therefore, the learnt representations of messages belonging to different events may concentrate in one direction and results in a certain dimension "waste". As shown in Fig.~\ref{fig_model}(a), to fully leverage the latent space and maximally differentiate the known events, we force the learnt features of events in different classes to be scattered in different directions by adding an orthogonal constraint. Specifically, we build a target pairwise similarity matrix based on the ground-truth events labels and demand the cosine similarity of the learnt message representations be close to it. Suppose we have $N_b$ nodes in a batch, the target pairwise similarity matrix is $\textbf{P}\in \mathbb{R}^{N_b\times N_b}$. $\textbf{P}_{ij}$ is $1$ if message $m_i$ and message $m_j$ belong to the same event; otherwise, the value is $0$.
The cosine similarities of the learnt message representations are computed as $\boldsymbol{\overline{H}}\cdot \boldsymbol{\overline{H}}^T$, where $\boldsymbol{\overline{H}}\in \mathbb{R}^{N_b\times d}$ denotes the normalized message representations. The additional orthogonal loss is: 
\begin{equation}
    \mathcal{L}_{o} = \text{Sum}((\boldsymbol{P} - \boldsymbol{\overline{H}}\times\boldsymbol{\overline{H}^T})^2),
\end{equation}
where $\text{Sum}(\cdot)$ represents the sum of the elements in the matrix.

Obviously, with the orthogonal inter-class relation constraint, directions are also utilized to train the message representations, which further differentiates the events in the latent space. Besides, with the orthogonal constraint, the known events become strong reference bases for those unknown samples thus greatly prompts knowledge transfer. 

\subsection{Self-improving Fine-tuning of GNN Encoder}\label{model_finetune}
\subsubsection{Pseudo Pairwise Labels Generation}
Previous works~\cite{hsu2018multi,han2020automatically} often use the trained model to get the initial representations of unknown data and generate pseudo labels based on their similarities. However, for those newly emerging events, it is very challenging to obtain discriminative representations due to their absence during the training process. Therefore, labels learnt from capturing merely the local data similarities are often of low quality. 

As illustrated in Fig.~\ref{fig_model}(b), to overcome this problem, we propose to utilize the Reference Similarity Distribution (RSD) vectors to generate pseudo labels. The idea is to learn soft multilabel vectors (real-valued label likelihood vectors) for those unknown samples by computing their similarities to a set of known reference events. Specifically, assume that $\boldsymbol{\overline{R}} = \{\boldsymbol{\overline{r}}_{e1},\boldsymbol{\overline{r}}_{e2},...\boldsymbol{\overline{r}}_{eK}\}\in\mathbb{R}^{K\times d}$ denotes a matrix which stores
the normalized representations of $K$ known reference events, with each row denoting a reference event. For example, $\boldsymbol{\overline{r}}_{ek}\in \mathbb{R}^d$ is the normalized cluster center of the $k$-th event. Suppose the normalized representation of a message sample $m_i$ from unknown data is denoted as $\boldsymbol{\overline{h}}_{m_i}$, its RSD vector $\boldsymbol{p}_{m_i}\in\mathbb{R}^K$ is calculated as:
\begin{equation}
    \boldsymbol{p}_{m_i} =\text{Softmax}(\boldsymbol{\overline{h}_{m_i}}\cdot \boldsymbol{\overline{R}}^T).
\end{equation}

The RSD vector, $\boldsymbol{p}_{m_i}$, captures the global similarities between $m_i$ and those known events. It contains much more knowledge compared to $\boldsymbol{\overline{h}_{m_i}}$, the originally learnt representation and thus has stronger discriminative power. Meanwhile, it is worth noting that those reference events are mutually discriminated from each other under the orthogonal constraint. This guarantees the effectiveness of them as strong reference bases. 

We use the cosine similarity between the RSD vectors of two messages to measure their consistency and generate the pseudo label for the pair. Suppose there is a pair of messages $(m_i,m_j)$, the consistency value $C(m_i,m_j)$ is calculated as:
\begin{equation}
    C(m_i,m_j)=\frac{\boldsymbol{p}_{m_i}\cdot(\boldsymbol{p}_{m_j})^T}{||\boldsymbol{p}_{m_i}||_2\cdot||(\boldsymbol{p}_{m_j})^T||_2},
\end{equation}
where $|| \cdot||_2$ denotes the $\ell_2$ norm.
 If $C(m_i,m_j)>0.5$, we set the pseudo label to $1$ (positive); otherwise, we set it to $0$ (negative).

\subsubsection{Pseudo Pairwise Labels Selection}\label{model_selection}
According to the pseudo pairwise label generation strategy described above, we get candidate pseudo labels of all possible pairs in a batch. Using all of them to fine-tune the model is both inefficient and unnecessary, because there are unavoidably numerous misclassified labels. How to select a small number of informative samples to achieve effective and efficient event knowledge transfer becomes an important problem. Inspired by the active learning~\cite{wu2021redal} which uses diversity as an indicator of sampling, we actively select a small number of pairs which are most different from the known data. In this way, the generalization ability of the learned model can be significantly strengthened with a limited training cost. 

In the unknown data, there are some samples which belong to the known events and some which are previously unseen. 
We have observed from experiments (Sec.~\ref{exp_entropy}) that the RSD vectors of samples in new events have higher information entropy compared to those within the known events. That's because a RSD vector, in its essence, is a soft multilabel probability vector. For those messages that belong to the known events, their RSD vectors tend to be one-hot vectors whose information entropy values are close to $0$. However, for messages within the new events, their distributions may concentrate in the space between several events, and thus the entropy is relatively large.  
Based on this observation, we propose to exploit
the entropy to estimate the diversity of a message sample.
Here, diversity refers to the degree of difference from the known events used to train the backbone model. To ensure effective knowledge transfer, we focus on the samples that are different from the known events during the fine-tuning stage.
Therefore, for each message, we use the entropy of its RSD vector to determine the number of its pairing messages. Formally, for message $m_i$, whose RSD vector is $\boldsymbol{p}_{m_i}$, the entropy is:
\begin{equation}
    H(\boldsymbol{p}_{m_i})=-\sum_{j=1}^{K} \boldsymbol{p}_{m_ij} \log _{2}\left(\boldsymbol{p}_{m_ij}\right).
\end{equation}
We employ an unbalanced sampling strategy and split the messages into two groups based on their entropy values. In experiments, for half of messages with larger entropy values, we sample 20 negative pairs and 20 positive pairs for model fine-tuning. For the other half, we sample 10 negative and 10 positive pairs.

\subsubsection{Pseudo Pairwise Labels Quality Assessment}
The quality assessment of those selected pseudo labels is also important. Without proper assessment and selection, the noisy pseudo-labels will gradually undermine the model. We have observed from experiments (Sec.~\ref{exp_quality}) that the consistency values (i.e. the cosine similarities between the RSD vectors) of pairs with wrong pseudo-labels are closer to the dividing threshold $0.5$. Based on this observation, we simply exploit the consistency value to estimate the quality of the pseudo labels. For positive pairs whose values are larger than $0.5$, the quality is positively correlated with consistency. On the contrary, for negative pairs whose values are smaller than $0.5$, the quality is negatively correlated with consistency. For simplicity, we directly assign the consistency value, $C(\text{Pos})$, of those pseudo positive pairs (consistency $>0.5$) as the quality, and assign $1-C(\text{Neg})$ as the quality of those pseudo negative ones, where Pos and Neg stand for a positive or negative message pair.

\subsubsection{Quality-guided Optimization}
To make the self-training (fine-tuning) process focus more on pair samples that the model is more confident on, we re-weight the importance of the pairs based on their quality. The quality-guided pairwise contrastive loss becomes:
\begin{equation}\label{equ_adj_pairloss}
	\begin{split} 
    \mathcal{L}_{qp}=\sum_{(m_{i}, m_{i}+)\in\{Pos\}
    \atop
    (m_j, m_{j-})\in\{Neg\} } (C(m_i,m_i+)+1-C(m_j,m_j-))\cdot
    \\\max \{\mathcal{D}(\boldsymbol{h}_{m_{i}}, \boldsymbol{h}_{m_{i}+})-\mathcal{D}(\boldsymbol{h}_{m_{j}}, \boldsymbol{h}_{m_{j}-})+a, 0\}.
    \end{split}
\end{equation}
During the fine-tuning process, we leverage the quality-guided pairwise contrastive loss computed from those selected pair samples to update the model. 

After model fine-tuning, we utilize the updated model to get the representations of all the unknown samples and adopt a specific clustering algorithm to output a set of social events. As for the clustering method, we can select distance-based clustering algorithms such as $K$-means or density-based ones such as DBSCAN~\cite{ester1996density}. Note that DBSCAN does not require specifying the total number of classes and thus is more suitable for open set social event detection.

\section{Experiments}
\subsection{Experimental Setup}
\subsubsection{Datasets}
We evaluate QSGNN on two large publicly available social event datasets: Events2012~\cite{mcminn2013building} and Events2018~\cite{mazoyer2020french}. We crawl the tweets via the Twitter API based on the provided IDs. After filtering out unavailable tweets, Events2012 contains 68,841 annotated tweets belonging to 503 event classes, and spreads over a period of 4 weeks. As for Events2018, it has 64,516 labeled tweets belonging to 257 event classes within 4 weeks.

\subsubsection{Baselines}
We compare QSGNN to both non-GNN-based methods and GNN-based methods. For the former, the baselines are: (1)\textbf{TwitterLDA}~\cite{zhao2011comparing} which is the first proposed topic model for Tweet data; (2) \textbf{Word2Vec}~\cite{mikolov2013}, which uses the average of the pre-trained Word2Vec embeddings of all words in the message as its representation; (3) \textbf{BERT}~\cite{devlin2018bert}, which uses the 768-d sentence embeddings of BERT as the message representations; (4) \textbf{EventX}~\cite{liu2020story}, which detects events based on community detection. For GNN-based methods, we select (5) \textbf{PP-GCN}~\cite{peng2019fine}, an offline fine-grained social event
detection method based on GCN. (6) \textbf{KPGNN}~\cite{cao2021knowledge} which leverages triplet loss to train GAT and gets message representations. 
% (7) \textbf{MVGAN}~\cite{cui2021mvgan} which utilizes GAT to learn representations of tweets from both semantic and temporal views.

\subsubsection{Implementation Details}
Our QSGNN is built on the PyTorch framework and on a machine equipped with seven NVIDIA GeForce RTX 3090 GPUs. As for the specific GNN encoder adopted in this work, we select a $2$-layer GAT network. We set the total number of heads to $4$, the hidden and output embedding dimensions to 32, the learning rate to 0.001, optimizer to Adam. In the pre-training stage, we set the training epochs to 15 with a patience of 5 for early stopping. In the fine-tuning stage, we set the training epochs to 3. Besides, we set the batch size to 2000 and the distance margin $a$ to 10. We repeat all experiments for 5 times and report the mean and standard deviation of the results. Some baselines (e.g., TwitterLDA, Bert) require the number of total event classes to be pre-defined. Thus, for a fair comparison, we apply the $K$-means clustering method after obtaining the message representations of all the 
models and set the total number of classes to the number of ground-truth classes. When applied in the real world, the $K$-means method can be replaced by DBSCAN, which does not require a pre-defined class number.

\subsubsection{Evaluation Metrics}
The performances of models are evaluated by two widely used clustering metrics: normalized mutual information (NMI) and adjusted mutual information (AMI). By measuring the amount of information from the distribution of the predictions, NMI has
been broadly adopted in event detection method evaluations. Meanwhile, considering NMI is not adjusted for chance, we also select AMI.

\begin{figure*}
\centering
\includegraphics[width=0.85\textwidth]{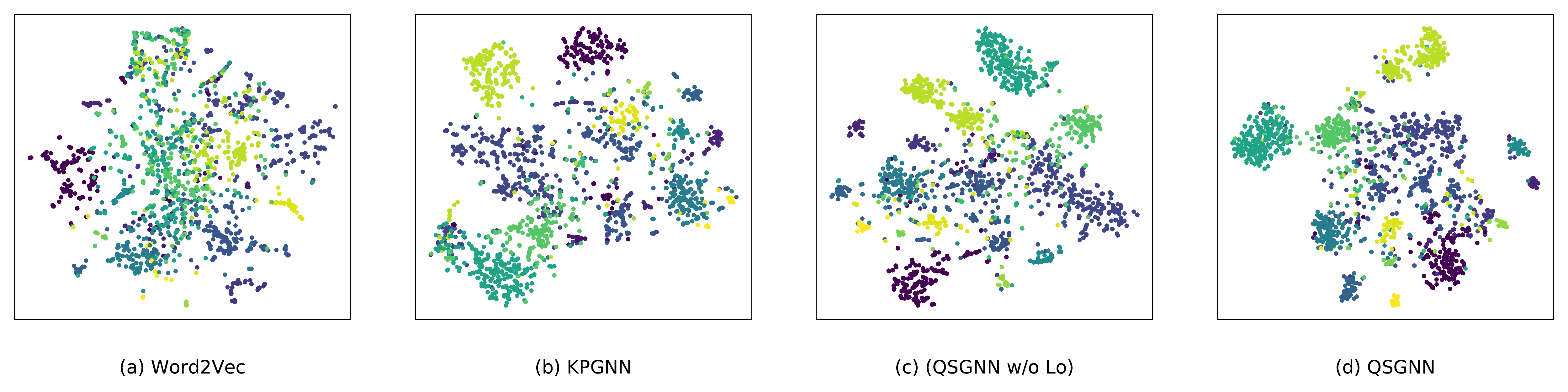}
\caption{Visualization on Events2018. Each node denotes a message and each color denotes an event class. } \label{fig_visualization}
\end{figure*}

\subsection{Evaluation on the Closed Set}
Recall that our model contains two stages: the supervised pre-training stage and the self-improving fine-tuning stage. In the pre-training stage, we assume the event labels are all available and utilize those known data in the initial block to train the backbone GNN encoder. While in the fine-tuning stage, messages from the incoming blocks are assumed to be unknown. To validate the effectiveness of our novel supervised pairwise contrastive learning method with the orthogonal constraint, we compare the performances in the closed set situation in which the training set, validation set and the test set share the same events.

Specifically, for both Events2012 and Events2018 datasets, we use the data of the first week to form the initial message block $M_0$. We randomly sample 20\% of the initial block for testing, 10\%
for validation, and use the rest 70\% for training.

\begin{table}
\caption{Evaluation on the Closed Set.}
\centering
\renewcommand\arraystretch{0.8}
\setlength{\tabcolsep}{1.6mm}
\begin{tabular}{ c |cc| cc}
\toprule
\multirow{2}{*}{Methods} &\multicolumn{2}{c|}{$M_0$ in Events2012} &\multicolumn{2}{c}{$M_0$ in Events2018}  \\
\cline{2-5}
\multirow{2}{*}{}& NMI & AMI  &NMI &AMI \\
\midrule
TwitterLDA~\cite{zhao2011comparing} &.26±.00 &.17±.00 &.22±.00 &.16±.00    \\
Word2Vec~\cite{mikolov2013} &.47±.00 &.21±.00& .24±.00 &.20±.00   \\
BERT~\cite{devlin2018bert} &.63±.01 &.44±.00 &.42±.00 &.34±.00   \\
EventX~\cite{liu2020story}   &.68±.00 &.29±.00 &.57±.00 &.56±.00    \\
PP-GCN~\cite{peng2019fine}   &.70±.02  &.56±.01  &.60±.01  &.49±.02      \\
KPGNN~\cite{cao2021knowledge}    & .76±.02 &.64±.02 &.66±.03 &.60±.02  \\
% MVGAN~\cite{cao2021knowledge}    & .63±.01 &.59±.01 &.27±.03 &.75±.02 \\
\midrule
QSGNN w/o $\mathcal{L}_{o}$  & .77±.00 & .65±.00  &.68±.02  &.61±.01\\ 
QSGNN  & \textbf{.79±.01} & \textbf{.68±.01}  &\textbf{.71±.02} &\textbf{.64±.02}\\ 
\hline
promotion &$\uparrow  3\%$ &$\uparrow 4\%$ &$\uparrow 5\%$ &$\uparrow 4\%$ \\
\bottomrule      
\end{tabular}  
\label{offevaluation}
\end{table}

\begin{table*}
\caption{Open set evaluation on the Events2012. }
\centering
\renewcommand\arraystretch{0.8}
\setlength{\tabcolsep}{0.5mm}
\begin{tabular}{c|cc|cc|cc|cc|cc|cc|cc}
\hline
Blocks &\multicolumn{2}{c|}{$M_1$} &\multicolumn{2}{c|}{$M_2$} &\multicolumn{2}{c|}{$M_3$} &\multicolumn{2}{c|}{$M_4$} &\multicolumn{2}{c|}{$M_5$} &\multicolumn{2}{c|}{$M_6$} &\multicolumn{2}{c}{$M_7$} \\ 
\hline
Metrics &NMI  &AMI &NMI  &AMI &NMI  &AMI &NMI  &AMI &NMI  &AMI &NMI  &AMI &NMI  &AMI \\
\hline
TwitterLDA &.11±.00&.08±.00& .27±.01&.20±.01& .28±.00&.22±.01& .25±.00&.17±.00& .26±.00&.21±.00& .32±.00&.20±.00& .18±.01&.12±.01\\
Word2Vec &.19±.00&.08±.00& .50±.00&.41±.00& .39±.00&.31±.00& .34±.00&.24±.00& .41±.00&.33±.00& .53±.00&.40±.00& .25±.00&.13±.00  \\
BERT &.36±.00&.34±.00& .78±.00&.76±.00& .75±.00&.73±.00& .60±.00&.55±.00& .72±.00&71±.00& .78±.00&.74±.00& .54±.00&.50±.00  \\
EventX &.36±.00&06±.00 &.68±.00&29±.00 &.63±.00&.18±.00 &.63±.00&19±.00 &.59±.00&14±.00 &.70±.00&27±.00 &.51±.00&13±.00  \\
PP-GCN &.23±.00&.21±.00 &.57±.02&.55±.02 &.55±.01&.52±.01 &.46±.01&.42±.01 &.48±.01&.46±.01 &.57±.01&52±.02 &.37±.00&.34±.00 \\
KPGNN &.39±.00&.37±.00  &.79±.01&.78±.01 &.76±.00&.74±.00  &.67±.00&.64±.01 &.73±.01&.71±.01 &.82±.01&.79±.01 &.55±.01&.51±.01  \\
% MVGAN  &.22±.00  &.22±.00 &.25±.00  &.28±.00  &.48±.00 &.33±.00  &.35±.00 &.37±.00 &.22±.00  &.22±.00 &.25±.00  &.28±.00  &.48±.00 &.33±.00  \\
\hline

QSGNN &\underline{\textbf{.43±.01}}  &\underline{\textbf{.41±.02}}  &\underline{\textbf{.81±.02}}    &\underline{\textbf{.80±.01}}   &\underline{\textbf{.78±.01}}  &\underline{\textbf{.76±.01}} &\underline{\textbf{.71±.02}} &\underline{\textbf{.68±.01}} &\underline{\textbf{.75±.00}}  &\underline{\textbf{.73±.00}} &\underline{\textbf{.83±.01}}  &\underline{\textbf{.80±.01}}  &\underline{\textbf{.57±.01}} &\underline{\textbf{.54±.00}} \\
\hline
promotion &$\uparrow  4\%$ &$\uparrow  4\%$ &$\uparrow  2\%$ &$\uparrow  2\%$ &$\uparrow  2\%$ &$\uparrow  2\%$ &$\uparrow  3\%$ &$\uparrow  3\%$&$\uparrow  2\%$ &$\uparrow  2\%$ &$\uparrow 1\%$ &$\uparrow 1\%$ &$\uparrow  2\%$ &$\uparrow  3\%$ \\
\hline
\hline
Blocks &\multicolumn{2}{c|}{$M_{8}$} &\multicolumn{2}{c|}{$M_{9}$} &\multicolumn{2}{c|}{$M_{10}$} &\multicolumn{2}{c|}{$M_{11}$} &\multicolumn{2}{c|}{$M_{12}$} &\multicolumn{2}{c|}{$M_{13}$} &\multicolumn{2}{c}{$M_{14}$} \\ 
\hline
Metrics &NMI  &AMI &NMI  &AMI &NMI  &AMI &NMI  &AMI &NMI  &AMI &NMI  &AMI &NMI  &AMI \\
\hline
TwitterLDA &.37±.01&24±.01& .34±.00&.24±.00& .44±.01&.36±.01&  .33±.01&.25±.01& .22±.01&.16±.01& .27±.00&.19±.00 &.21±.00&.15±.01  \\
Word2Vec&.46±.00&.33±.00& .35±.00&.24±.00& .51±.00&.39±.00& .37±.00&.26±.00& .30±.00&.23±.00& .37±.00&.23±.00& .36±.00&.26±.00  \\
BERT &.79±.00&.75±.00& .70±.00&.66±.00& .74±.00&.70±.00& .68±.00&.65±.00& .59±.00&.56±.00& .63±.00&.59±.00& .64±.00&.61±.00 \\
%DeepWalk &.6071  & .6047 &.6860  &.5759  &.6077 &.6610  &.7403 %&.6178  \\
EventX   &71±.00&.21±.00& .67±.00&.19±.00& .68±.00&.24±.00& .65±.00&.24±.00& .61±.00&.16±.00& .58±.00&.16±.00& .57±.00&.14±.00  \\
PP-GCN &.55±.02&.49±.02& .51±.02&.46±.02& .55±.02&.51±.02& .50±.01&.46±.02& .45±.01&.42±.01& .47±.01&.43±.01& .44±.01&.41±.01  \\
KPGNN &\underline{\textbf{.80±.00}}&\underline{\textbf{76±.01}}& .74±.02&.71±.02& .80±.01&78±.01&.74±.01&.71±.01& .68±.01&66±.01 &\underline{\textbf{.69±.01}}&\underline{\textbf{.67±.01}}& \underline{\textbf{.69±.00}}&.65±.00  \\
% MVGAN  &.22±.00  &.22±.00 &.25±.00  &.28±.00  &.48±.00 &.33±.00  &.35±.00 &.37±.00 &.22±.00  &.22±.00 &.25±.00  &.28±.00  &.48±.00 &.33±.00  \\
\hline
QSGNN &.79±.01  &.75±.01  &\underline{\textbf{.77±.02}}   &\underline{\textbf{.75±.02}}   &\underline{\textbf{.82±.02}}  &\underline{\textbf{.80±.03}}   &\underline{\textbf{.75±.01}} &\underline{\textbf{.72±.01}} &\underline{\textbf{.70±.00}}  &\underline{\textbf{.68±.00}} &.68±.02  &.66±.01  &.68±.01 &\underline{\textbf{.66±.01}} \\
\hline
promotion &$\downarrow  1\%$ &$\downarrow  1\%$ &$\uparrow  3\%$ &$\uparrow  4\%$ &$\uparrow  2\%$ &$\uparrow  2\%$ &$\uparrow  1\%$ &$\uparrow  1\%$&$\uparrow  2\%$ &$\uparrow  2\%$ &$\downarrow  1\%$ &$\downarrow  1\%$ &$\downarrow  1\%$ &$\uparrow  1\%$ \\
\hline
\hline
Blocks  &\multicolumn{2}{c|}{$M_{15}$} &\multicolumn{2}{c|}{$M_{16}$} &\multicolumn{2}{c|}{$M_{17}$} &\multicolumn{2}{c|}{$M_{18}$} &\multicolumn{2}{c|}{$M_{19}$} &\multicolumn{2}{c|}{$M_{20}$}&\multicolumn{2}{c}{$M_{21}$}\\ 
\hline
Metrics &NMI  &AMI &NMI  &AMI &NMI  &AMI &NMI  &AMI &NMI  &AMI &NMI  &AMI &NMI  &AMI \\
\hline
TwitterLDA &.21±.00&.13±.00& .35±.01&.27±.01& .19±.00&.13±.00& .18±.00&.12±.00& .29±.01&.22±.00& .35±.00&.23±.00& .19±.00&.13±.00  \\
Word2Vec&.27±.00&.15±.00& .49±.00&.36±.00& .33±.00&.24±.00& .29±.00&.21±.00& .37±.00&.28±.00& .38±.00&.24±.00& .31±.00&.21±.00  \\
BERT &54±.00&.50±.00 &.75±.00&.72±.00 &.63±.00&.60±.00 &.57±.00&.53±.00 &.66±.00&.63±.00 &.68±.00&.62±.00 &.59±.00&57±.00  \\
%DeepWalk &.6071  & .6047 &.6860  &.5759  &.6077 &.6610  &.7403 %&.6178  \\
EventX  &.49±.00&07±.00& .62±.00&.19±.00& .58±.00&.18±.00& .59±.00&.16±.00& .60±.00&.16±.00& .67±.00&.18±.00 &.53±.00&.10±.00  \\
PP-GCN &.39±.01&.35±.01& .55±.01&.52±.01& .48±.00&.45±.00& .47±.01&.45±.01& .51±.02&.48±.02& .51±.01&.45±.02& .41±.02&.38±.02  \\
KPGNN &.58±.00&.54±.00& \underline{\textbf{.79±.01}}&\underline{\textbf{.77±.01}}& .70±.01&.68±.01& .68±.02&.66±.02& \underline{\textbf{.73±.01}}&\underline{\textbf{.71±.01}}& .72±.02&.68±.02& \underline{\textbf{.60±.00}}&.57±.00  \\
% MVGAN  &.22±.00  &.22±.00 &.25±.00  &.28±.00  &.48±.00 &.33±.00  &.35±.00 &.37±.00 &.22±.00  &.22±.00 &.25±.00  &.28±.00  &.48±.00 &.33±.00  \\
\hline
QSGNN &\underline{\textbf{.59±.01}} &\underline{\textbf{.55±.01}}  &.78±.01    &.76±.02  &\underline{\textbf{.71±.01}}  &\underline{\textbf{.69±.01}}   &\underline{\textbf{.70±.01}} &\underline{\textbf{.68±.01}} &\underline{\textbf{.73±.00}}  &.70±.01 &\underline{\textbf{.73±.02}}  &\underline{\textbf{.69±.02}}  &\underline{\textbf{.61±.01}} &\underline{\textbf{.58±.00}} \\
\hline
promotion &$\uparrow  1\%$ &$\uparrow 1\%$ &$\downarrow  1\%$ &$\downarrow  1\%$ &$\uparrow  1\%$ &$\uparrow  1\%$&$\uparrow  2\%$ &$\uparrow  2\%$&- &$\downarrow  1\%$ &$\uparrow  1\%$ &$\uparrow  1\%$ &$\uparrow  1\%$ &$\uparrow  1\%$ \\
\hline

\end{tabular}
\label{table_2012results}
\end{table*}

\subsubsection{Comparison with the state-of-the-arts}
As shown in Table~\ref{offevaluation}, QSGNN yields the best results. That's because it fully utilizes space distance as well as direction information to distinguish different events. Compared to KPGNN, which only uses distance information to learn event representations, QSGNN achieves 3\% and 5\% performance gains in NMI on Events2012 and Events2018, respectively. It is worth noting that, even without the direction information, i.e., the orthogonal inter-class constraint, (QSGNN w/o $\mathcal{L}_{o}$) still works better than KPGNN. That is due to the more strict distance constraint. The triplet loss adopted in KPGNN only requires the intra-class distance to be smaller than the inter-class distance of the same anchor. However, the pairwise loss proposed in QSGNN demands the inter-class distance to be smaller than the minimum inter-class distance. Therefore, the intra-class representations learnt by (QSGNN w/o $\mathcal{L}_{o}$) are more distinguishable from the inter-class ones. Besides, we also notice that GNN-based methods (i.e., PP-GCN, KPGNN and QSGNN) perform much better than general message representation learning methods (i.e., Word2Vec and BERT) and word distribution methods like TwitterLDA. For example, QSGNN gets a large improvement (53\%) in NMI compared to TwitterLDA in Events2012. We owe this contribution to the effectiveness of GNN-based methods in exploring the graph structure contained in the social network.

 \begin{table*}
\caption{Open set evaluation on the Events2018. }
\centering
\renewcommand\arraystretch{0.8}
\setlength{\tabcolsep}{0.3mm}
\begin{tabular}{c|cc|cc|cc|cc|cc|cc|cc|cc}
\hline
Blocks &\multicolumn{2}{c|}{$M_1$} &\multicolumn{2}{c|}{$M_2$} &\multicolumn{2}{c|}{$M_3$} &\multicolumn{2}{c|}{$M_4$} &\multicolumn{2}{c|}{$M_5$} &\multicolumn{2}{c|}{$M_6$} &\multicolumn{2}{c|}{$M_7$} &\multicolumn{2}{c}{$M_8$}\\ 
\hline
Metrics &NMI  &AMI &NMI  &AMI &NMI  &AMI &NMI  &AMI &NMI  &AMI &NMI  &AMI &NMI  &AMI &NMI  &AMI\\
\hline
TwitterLDA &.20±.00&19±.00  &.09±.00&06±.00 &.13±.00&.11±.00  &.10±.00&.08±.00 &.24±.00&.20±.00  &.22±.00&.19±.00  &.12±.00&.10±.00  &.24±.00&.20±.00 \\
Word2Vec &.22±.00&21±.00   &.22±.00&21±.00  &.25±.00&23±.00   &.28±.00&27±.00   &.48±.00&46±.00  &.33±.00&31±.00   &.35±.00&.33±.00  &.37±.00&34±.00  \\
BERT &.32±.00&.28±.00   &.32±.00&.31±.00 &.31±.00&.32±.00  &.33±.00&.30±.00  &.47±.00&.44±.00 &.36±.00&.33±.00  &.41±.00&.36±.00 &.44±.00&.38±.00\\
%DeepWalk &.6071  & .6047 &.6860  &.5759  &.6077 &.6610  &.7403 %&.6178  \\
EventX   &.34±.00&.11±.00   &.37±.00&.12±.00 &.37±.00&.11±.00  &.39±.00&.14±.00   &.53±.00&.24±.00 &.44±.00&.15±.00   &.41±.00&.12±.00 &.54±.00&.21±.00\\
PP-GCN &.49±.01&48±.00  &.45±.00&44±.02 &.56±.03&.55±.03  &.54±.03&.54±.04 &.54±.02&.53±.02 &.52±.02&.50±.03 &.56±.04&.55±.04  &.56±.03&.55±.02 \\
KPGNN &.54±.01&.54±.01  &.56±.02&.55±.01  &.52±.03&.55±.02 &.55±.01&.55±.01  & .58±.02&57±.01 &.59±.03&.57±.02  &.63±.02&61±.02  &\underline{\textbf{.58±.02}}&\underline{\textbf{.57±.02}}  \\
% MVGAN  &.54±.01  &.56±.02  &.52±.03  & .55±.01  & .58±.02  &.59±.03  &.63±.02 &.58±.02 &.22±.00  &.22±.00 &.25±.00  &.28±.00  &.48±.00 &.33±.00  &.35±.00 &.37±.00 \\
\hline
QSGNN &\underline{\textbf{.57±.01}}  &\underline{\textbf{.56±.01}} &\underline{\textbf{.58±.01}} &\underline{\textbf{.57±.01}}   &\underline{\textbf{.57±.01}}  &\underline{\textbf{.56±.02}} 
&\underline{\textbf{.58±.03}}  &\underline{\textbf{.57±.03}}
&\underline{\textbf{.61±.02}} &\underline{\textbf{.59±.01}}  &\underline{\textbf{.60±.01}} &\underline{\textbf{.59±.01}} &\underline{\textbf{.64±.01}}  &\underline{\textbf{.63±.01}} &.57±.02  &.55±.02  \\

\hline
promotion &$\uparrow  2\%$ &$\uparrow  1\%$ &$\uparrow  2\%$ &$\uparrow  2\%$ &$\uparrow  1\%$ &$\uparrow  1\%$ &$\uparrow  3\%$ &$\uparrow  2\%$&$\uparrow  3\%$ &$\uparrow  2\%$ &$\uparrow  1\%$ &$\uparrow  2\%$ &$\uparrow  1\%$ &$\uparrow  2\%$ &$\downarrow  2\%$ &$\downarrow  2\%$\\
\hline
\hline
Blocks &\multicolumn{2}{c|}{$M_9$} &\multicolumn{2}{c|}{$M_{10}$} &\multicolumn{2}{c|}{$M_{11}$} &\multicolumn{2}{c|}{$M_{12}$} &\multicolumn{2}{c|}{$M_{13}$} &\multicolumn{2}{c|}{$M_{14}$} &\multicolumn{2}{c|}{$M_{15}$} &\multicolumn{2}{c}{$M_{16}$}\\ 
\hline
Metrics &NMI  &AMI &NMI  &AMI &NMI  &AMI &NMI  &AMI &NMI  &AMI &NMI  &AMI &NMI  &AMI &NMI  &AMI\\
\hline
TwitterLDA &.16±.00&.12±.00  &.17±.00&.11±.00 &.22±.00&.18±.00 &.28±.00&.25±.00  &.19±.00&.17±.00 &.24±.00&.21±.00  &.33±.00&.30±.00 &.07±.00&.02±.0 \\
Word2Vec &.33±.00&.30±.00  &.46±.00&.42±.00 &.41±.00&38±.00  &.40±.00&.37±.00   &.22±.00&20±.00 &.36±.00&34±.00  &.41±.00&38±.00 &.28±.00&.25±.00 \\
BERT &.38±.00&.28±.00  &.42±.00&.35±.00 &.45±.00&.34±.00  &.48±.00&.44±.00 &.31±.00&.26±.00 &.43±.00&.40±.00  &.39±.00&.39±.00 &.34±.00&.27±.00\\
%DeepWalk &.6071  & .6047 &.6860  &.5759  &.6077 &.6610  &.7403 %&.6178  \\
EventX   &.45±.00&.16±.00   &.52±.00&.19±.00  &.48±.00&.18±.00   &.51±.00&.20±.00   &.44±.00&.15±.00  &.52±.00&.22±.00  &.49±.00&.22±.00 &.39±.00&.10±.00\\
PP-GCN &\underline{\textbf{.54±.02}}&\underline{\textbf{.48±.03}} &.56±.06&.55±.04 &.59±.03&.57±.02  &.60±.02&.58±.02 &\underline{\textbf{.61±.01}}&.59±.02 &.60±.02&.59±.01 &.57±.03&.55±.03  &\underline{\textbf{.53±.02}}&\underline{\textbf{.52±.02}}  \\
KPGNN &.48±.02&.46±.02  &.57±.01&.56±.02  &.54±.01&.53±.01 &.55±.04&.56±.02 &.60±.02&\underline{\textbf{.60±.02}}  &.66±.01&.65±.00  &.60±.01&.58±.02  &.52±.02&.50±.01\\
% MVGAN  &.54±.01  &.56±.02  &.52±.03  & .55±.01  & .58±.02  &.59±.03  &.63±.02 &.58±.02 &.22±.00  &.22±.00 &.25±.00  &.28±.00  &.48±.00 &.33±.00  &.35±.00 &.37±.00 \\
\hline
QSGNN  &.52±.02 &.46±.02 &\underline{\textbf{.60±.01}} &\underline{\textbf{.58±.01}} &\underline{\textbf{.60±.01}} &\underline{\textbf{.59±.02}}  &\underline{\textbf{.61±.02}}   &\underline{\textbf{.59±.02}} &.59±.04  &.58±.03 &\underline{\textbf{.68±.02}}  &\underline{\textbf{.67±.02}}  &\underline{\textbf{.63±.02}} &\underline{\textbf{.61±.00}}  &.51±.03 &.50±.03 \\

\hline
promotion &$\downarrow  2\%$ &$\downarrow  2\%$ &$\uparrow  3\%$ &$\uparrow  2\%$ &$\uparrow  1\%$ &$\uparrow  2\%$ &$\uparrow 1\%$ &$\uparrow  1\%$ &$\downarrow  2\%$ &$\downarrow  2\%$ &$\uparrow  2\%$ &$\uparrow  2\%$ &$\uparrow  3\%$ &$\uparrow  3\%$ &$\downarrow  2\%$ &$\downarrow  2\%$\\
\hline
\end{tabular}
\label{table_2018results}
\end{table*}

\subsubsection{Visualization}
For a more intuitive comparison and to further show the effectiveness of our proposed QSGNN, we conduct visualization on Events2018 by plotting the representations of the test set using t-SNE. The results are illustrated in Fig.~\ref{fig_visualization}.
Obviously, GNN-based methods which capture both semantics and internal structure information are capable to learn more distinguishable representations compared to Word2Vec. Meanwhile, the observation that intra-class representations learnt by (QSGNN w/o $\mathcal{L}_{o}$) gather more closely compared to KPGNN verifies the effectiveness of our novel pairwise contrastive learning loss. Furthermore, the more concentrated intra-class distribution in QSGNN compared to (QSGNN w/o $\mathcal{L}_{o}$) demonstrates the effects of adding orthogonal constraint during training.

\subsection{Evaluation on the Open Set}
In the self-improving fine-tuning stage, we assume the data is unknown. To achieve knowledge transferring and make the model adapt to the incoming new data, our model generates pseudo labels to update the initial model. However, for most baselines such as PP-GCN and KPGNN, it is necessary to continuously provide event data with labels to train the model for the new message blocks.
To compare with them, for those baselines which need supervision, we still follow the operation in the closed set - sampling 70\% for training, 10\% for validation and 20\% for testing. Note that our model is superior to those methods since it does not require external annotation which is labour-costly.

\begin{table}
\centering
\renewcommand\arraystretch{1}
\setlength{\tabcolsep}{0.6mm}
  \caption{Consistency values calculated from the initial representations $\boldsymbol{h}$ vs. from RSD vectors.}
  \label{table_rsd}
  \begin{tabular}{c|c|c|c|c}
    \toprule
    \multirow{3}{*}{label} &\multicolumn{2}{c|}{Events2012}&\multicolumn{2}{c}{Events2018}\\
    \cline{2-5}
    &Consistency  &Consistency  &Consistency  &Consistency \\
    &of $\boldsymbol{h}$&of RSD&of $\boldsymbol{h}$&of RSD\\
    \cline{1-5}
    Positive &0.7287 &0.7644 &0.6576 &0.8350\\
    Negative &0.1834 &0.1002 &0.3023 &0.3459\\
    \hline
    difference&0.5453 &0.6642&0.3553&0.4891\\
  \bottomrule
\end{tabular}
\end{table}

\subsubsection{Comparison with the state-of-the-arts}
We demonstrate the results in Table~\ref{table_2012results} and Table~\ref{table_2018results}. Generally, QSGNN outperforms the strongest baseline, KPGNN, in most message blocks (with 1\%-4\% performance gains). Note that the proposed QSGNN, unlike KPGNN and the other baselines, does not require ground-truth labels for continuous model training or updates. It is impressive that QSGNN, which gets fine-tuned only by the generated pseudo pairwise labels, performs even better than those supervised baselines. This varifies the superiority and effectiveness of our model in extending knowledge from known to unknown.

% \begin{figure}
% \centering
% \includegraphics[width=0.46\textwidth]{image/rsd.pdf}
% \caption{Consistency values calculated by the initial representations vs RSD vectors. } \label{fig_rsd}
% \end{figure}

\subsubsection{The consistency values of positive and negative pairs}
To demonstrate the superiority of utilizing the learnt reference distribution similarity to generate pseudo pairwise labels, we record the average consistency values (cosine similarities) of real positive pairs and real negative pairs calculated from the initially learnt representations and the RSD vectors, respectively, in Table~\ref{table_rsd}. Obviously, when calculated from the RSD vectors, the consistency values of the positive pairs and the negative pairs differ more. Thus we can give more accurate judgement to the positive/negative pairs. Hence, the
proposed RSD vector is helpful.

\subsubsection{Using entropy to measure the diversity of events}\label{exp_entropy}
\begin{figure}
\centering\includegraphics[width=0.44\textwidth]{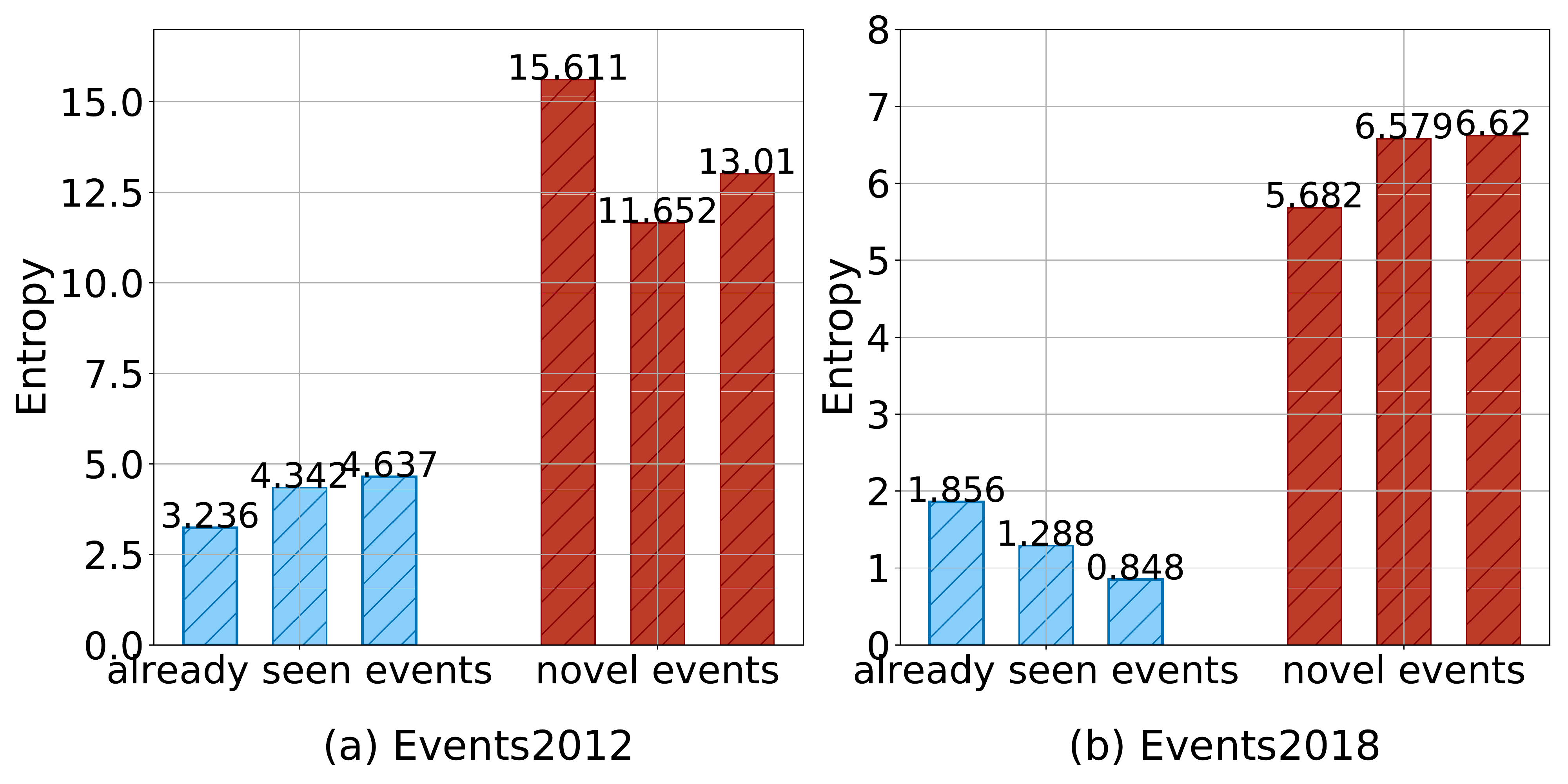}
\caption{The diversity of events measured by entropy. } \label{fig_entropy}
\end{figure}

We plot the average entropy of $3$ randomly selected events which are previously seen in the pre-training stage and $3$ randomly selected novel events which have not appeared in the pre-training stage in Fig.~\ref{fig_entropy}. Apparently, those novel events have higher information entropy which means they are more different from the known data. This validates the effectiveness of using entropy to sample diverse events.

\begin{figure}
\centering
\includegraphics[width=0.46\textwidth]{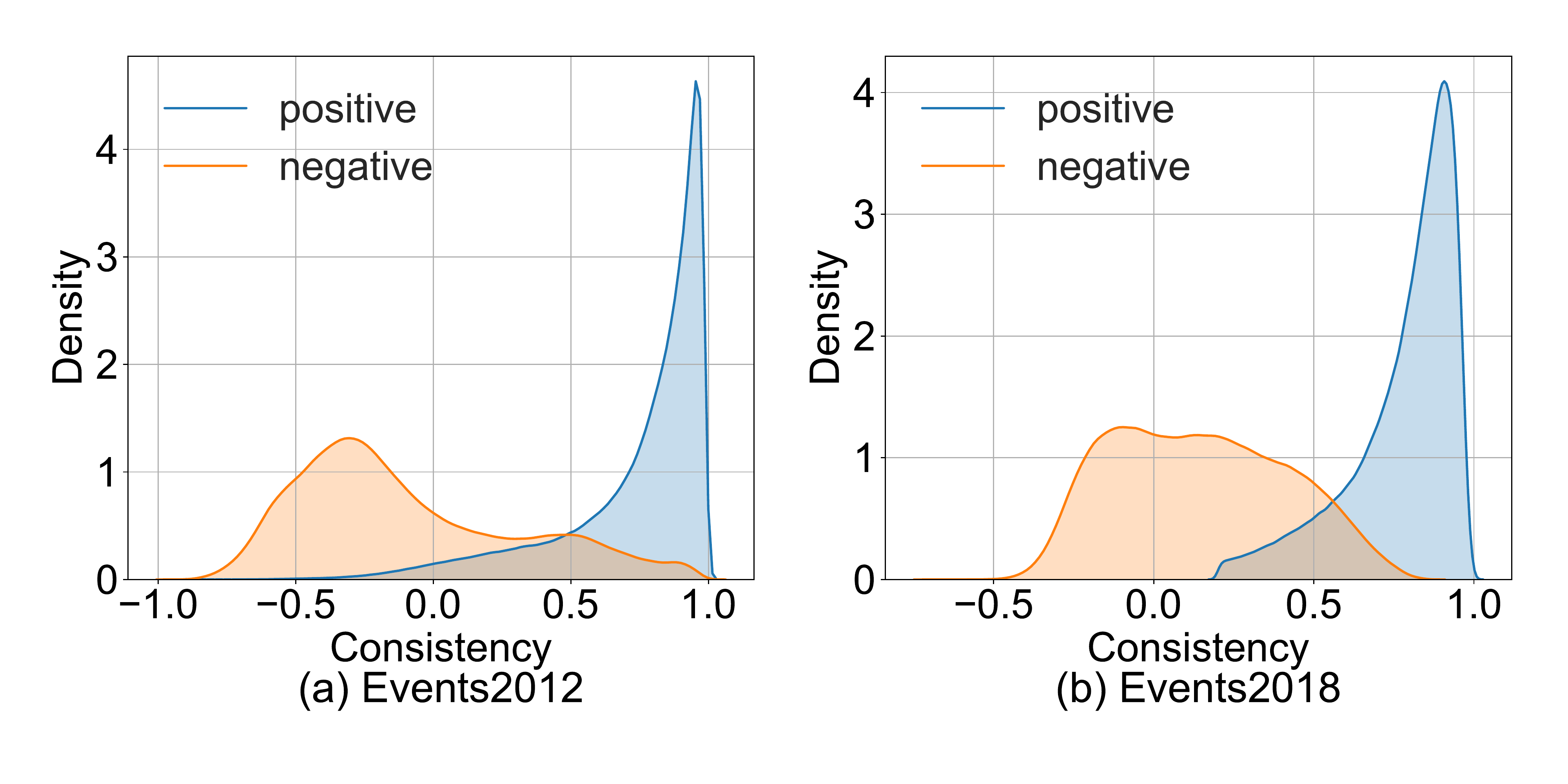}
\caption{The distribution of consistency values. } \label{fig_quality}
\end{figure}

\begin{figure}
\centering
\includegraphics[width=0.46\textwidth]{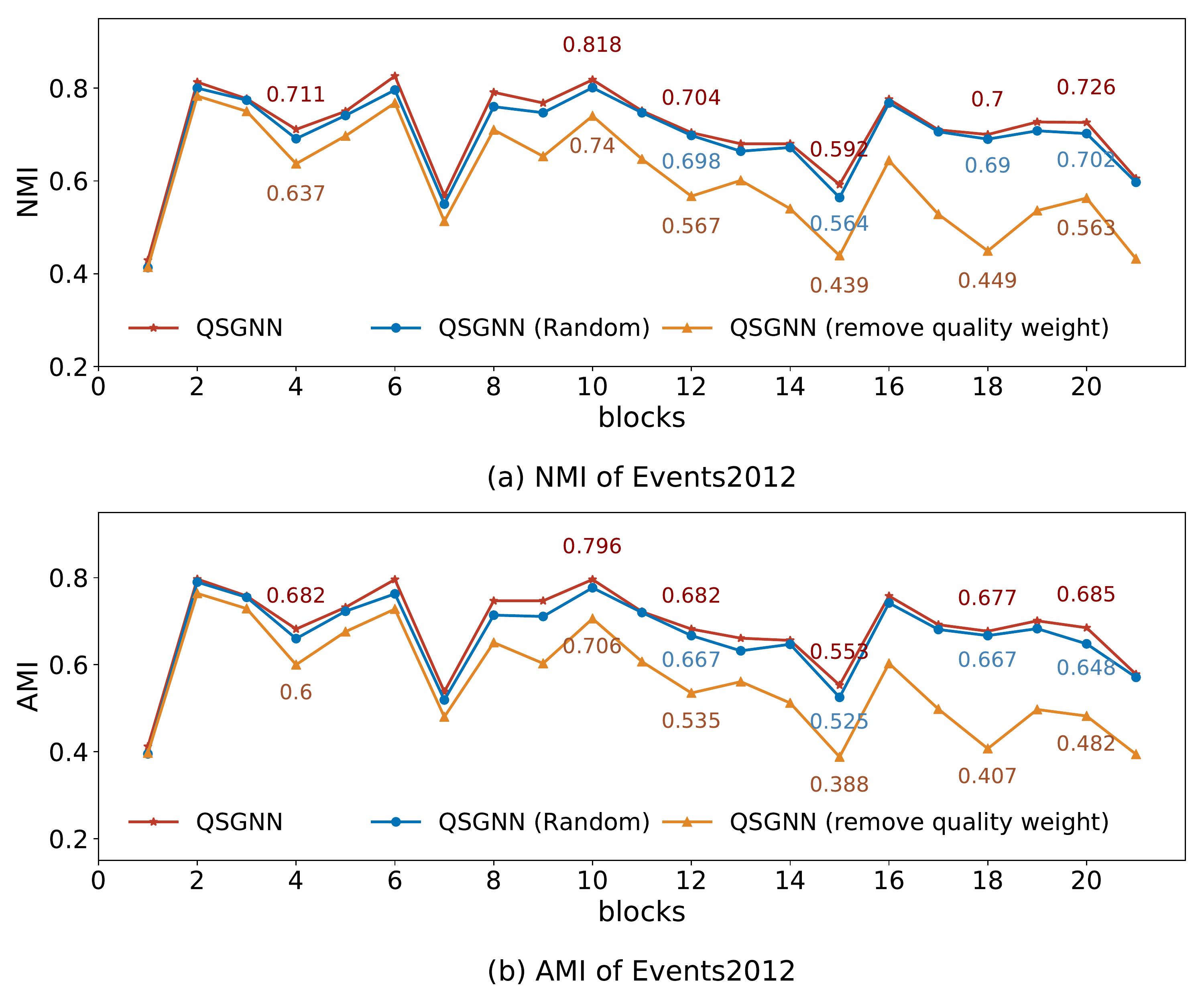}
\caption{Results with different selection strategy and loss. } \label{fig_ablation}
\end{figure}

\subsubsection{Using consistency value to measure pairwise label quality}~\label{exp_quality}
We plot the distributions of consistency values of real positive pairs and real negative pairs in Fig.~\ref{fig_quality}. When the consistency value $>0.5$, as the value increases, the percentage of positive pairs also increases. Similarly, when the consistency value $<0.5$, as the value decreases, the percentage of negative pairs gets higher. When the consistency is at a maximum or minimum, the qualities of the pseudo labels are highest. However, when the consistency value is close to the threshold ($0.5$), the positive and negative labels have the highest mixing ratio thus are unreliable. Fig.~\ref{fig_quality} demonstrates the relation between the consistency value and the label distribution to measure the label quality.

\subsubsection{Ablation study}
To validate the usefulness of (1) the selection strategy based on the diversity principle and (2) the quality-aware optimization, we perform an ablation study. 
For (1), we remove the unbalanced sampling strategy, as mentioned in Sec.~\ref{model_selection}, and 
adopt a $Random$ strategy to make a comparison. Specifically, we randomly select 15 negative and 15 positive pairs of each message. For (2), we remove the quality weight and adopt
$\mathcal{L}_{t}$ to compute the loss. We show the results in Fig.~\ref{fig_ablation}. When the unbalanced selection strategy is replaced by the $Random$ selection strategy, the performances drop slightly. This demonstrates the superiority of the diversity-based sample selection. By selecting more samples which are different from the known events, the model better adapts to the newly emerging data. Besides, from Fig.~\ref{fig_ablation}, we can see that when the quality-aware optimization is removed, the results have a significant decline. This validates the indispensability of the quality assessment. By measuring the quality of pseudo labels and adjusting their contributions to the loss, the fine-tuning is more reliable.

% \subsection{Parameter Analysis}
\section{Conclusion}
We have presented a quality-aware self-improving GNN framework to
tackle the challenging problem of open set social event detection. First, to make the best of those known events,
we extend the conventional triplet loss to a more strict pairwise loss with an orthogonal constraint to train the GNN encoder. Next, to generalize from known to unknown in an effective and reliable way, we propose to use the reference similarity distribution vectors for pseudo pairwise label generation, selection and quality assessment. Specifically, the selection strategy follows the principle of diversity and the quality is measured by consistency.
A quality-aware optimization strategy is proposed to resist the noise by re-weighting the contributions of different pseudo labels.
Experimental results illustrate that our model achieves state-of-the-art results in both closed set setting and open set setting.

% \begin{table*}
%   \caption{Some Typical Commands}
%   \label{tab:commands}
%   \begin{tabular}{ccl}
%     \toprule
%     Command &A Number & Comments\\
%     \midrule
%     \texttt{{\char'134}author} & 100& Author \\
%     \texttt{{\char'134}table}& 300 & For tables\\
%     \texttt{{\char'134}table*}& 400& For wider tables\\
%     \bottomrule
%   \end{tabular}
% \end{table*}

\begin{acks}
The authors of this paper were supported by the National Key R\&D Program of China through grant 2021YFB1714800, NSFC through grants U20B2053 and 62002007, S\&T Program of Hebei through grant 21340301D, Beijing Natural Science Foundation through grant 4222030, and the Fundamental Research Funds for the Central Universities. 
Philip S. Yu was supported by NSF under grants III-1763325, III-1909323, III-2106758, and SaTC-1930941.
Thanks for computing infrastructure provided by Huawei MindSpore platform.
For any correspondence, please refer to Hao Peng.
\end{acks}

%%
%% The next two lines define the bibliography style to be used, and
%% the bibliography file.
\bibliographystyle{ACM-Reference-Format}
\bibliography{sample-base}

%%
%% If your work has an appendix, this is the place to put it.
% \appendix

\end{document}